\documentstyle[12pt,epsfig,epsf]{article}
\begin{document}
\thispagestyle{empty}
\begin{center}
\LARGE
\textbf{Lattice Boltzmann simulations  of
segregating binary fluid mixtures in shear flow}
~\\
~\\
\vspace{1.cm}
\normalsize
A. Lamura$^{1,2}$ and G. Gonnella$^1$\\
\vspace{0.5cm}
$^1${\it Istituto Nazionale per la Fisica della Materia, Unit\`{a} di Bari} and\\ 
{\it Dipartimento di Fisica dell'Universit\`{a} di Bari} and\\
{\it Istituto Nazionale di Fisica Nucleare, Sezione di Bari\\
via Amendola 173, 70126 Bari, Italy}\\
\vskip .35 cm
$^2${\it Institut f\"{u}r Festk\"{o}rperforschung,
Forschungszentrum J\"{u}lich,}\\
{\it 52425 J\"{u}lich, Germany}\\
~\\
~\\
\end{center}
\vspace{1.cm}
\begin{abstract}
\noindent
We apply  lattice Boltzmann method to study the
phase separation of a
two-dimensional binary fluid mixture in shear flow.
The algorithm can simulate systems described by
the Navier-Stokes and  convection-diffusion equations.
We propose a new scheme for 
imposing the shear flow which has
the advantage of preserving mass and momentum conservation 
on the boundary walls without introducing  slip velocities.
Our main results concern  the presence of two typical lenght scales
in the phase separation process, corresponding  to  domains with two different
thicknesses. 
Our simulations at  low  viscosity
 confirm previous results only valid in the limit of infinite
viscosity.
\end{abstract}
\newpage
\addtolength{\baselineskip}{\baselineskip}
\par\noindent
\section{Introduction}
\par
When a fluid mixture is suddenly quenched from a disordered initial 
condition to a coexistence state
below the  critical temperature, 
the two fluids  segregate with domains growing with time
(see, e.g., Gunton {\it et al.} 1983). The morphology of these
domains  is 	strongly 
influenced by an applied flow.
In the case of a shear flow 
(for a review see Onuki 1997),  the domains 
 are observed to grow greatly
elongated in the flow direction
(see, e.g., Hashimoto {\it et al.} 1995).
The effects of the shear have been considerably investigated in the last years.
Several numerical simulations confirm
the anisotropic growth of the
 domains (Ohta {\it et al.} 1990, Rothman 1991, Olson \& Rothman 1995,
Wu {\it et al.} 1997, Padilla \& Toxvaerd 1997, Shou \& Chakrabarti 2000).
New recent theoretical results  by Corberi {\it et al.} (1998)-(1999)
show that, as an effect of the shear,
 the mixture contains 
 domains of  two different thicknesses and that the relative abundance
of these domains changes periodically with the logarithm of time. 
However in the papers of  Corberi {\it et al.} (1998)-(1999)
 phase separation occurred only for the effects of simple diffusion.
The main purpose of this paper is to see how the hydrodynamics
affects this  picture.

The presence of domains with different scales  is a 
very peculiar result for phase separating mixtures.
Indeed, in the general picture for the unsheared case, 
a single time-dependent length scale  $R(t)$, which measures the typical size
of domains, characterizes the behaviour of the mixture
(see Bray 1994). 
This  length grows with the 
power law $R(t) \sim t^{\alpha}$, 
where the value of the  growth exponent 
  $\alpha$ is strictly related to the physical mechanism
operating in the segregation process. 
For example  $\alpha=1/3$   in a pure diffusive regime.
Therefore the existence of two characteristic scales 
makes the phase separation process very interesting
from a theoretical point of view and also with relevant
practical implications.

Looking at the effects of hydrodynamics on this problem
is not a simple computational task.
Domains grow fast in the direction
of the flow  making 
finite-size effects quite soon   relevant in simulations
(for a recent review, see, Yeomans 1999).
In this paper we use a lattice Boltzmann scheme
 (Orlandini {\it et al.} 1995, Swift {\it et al.} 1996)
to simulate the 
convection-diffusion and Navier-Stokes equations for 
 a binary mixture. This method has been found very convenient
with respect to other numerical algorithms 
for  fluid mixtures (Chen \& Doolen 1998) 
since it allows to reach very large time scales.
In particular, 
an  advantage of the lattice Boltzmann scheme of
Orlandini {\it et al.} (1995) and Swift {\it et al.} (1996), 
is that  the correct fluid equilibrium 
can be  imposed by choosing an appropriate free energy. 
 Moreover, differently from other lattice gas methods,
it  allows to control in a separate way the 
different mechanisms of
transport tuning independently the fluid viscosity and the diffusivity.
The mentioned lattice Boltmann scheme 
 has been
used with success in several problems of phase separation 
of binary fluid mixtures without shear 
in two (Osborn {\it et al.} 1995)
and three dimensions (Kendon {\it et al.} 1999),
and also in the case of 
 complex fluids (Gonnella {\it et al.} 1997).

A preliminary important  part of our  work 
has been to  include  proper boundary conditions for a shear flow
in  lattice Boltzmann schemes. 
This problem, considered also  by
Wagner \& Yeomans (1998) and 
Cates {\it et al.} (1999), has been studied in many papers
on lattice Boltzmann  schemes for a single fluid
 (for a review, see, e. g., Chen \& Doolen 1998). 
We discuss previous procedures and propose an improved algorithm
which allows to strictly conserve  mass and momentum on the boundary walls
without introducing slip velocities.

The outline of the paper is as follows. In the next section we describe 
the fluid model
 and the lattice Boltzmann scheme. In section 3 we describe
 and compare
 different approaches for  implementing shear boundary conditions
in the numerical scheme. In section 4 
we give our results for the
 phase separation and
 finally we draw our conclusions.
\section{The lattice Boltzmann method}
\par
Our simulations are based on  a lattice Boltzmann scheme 
developed by Orlandini {\it et al.} (1995) and Swift {\it et al.} (1996).
In this scheme the 
choice of a  free energy determines all the thermodinamic properties
of the fluid. We start from the description of the free energy
of the fluid mixture; then we show how the lattice Boltzmann
equations have been implemented.

{\it Equilibrium description of the mixture.} The  
free-energy functional generally used in phase separation studies 
of binary mixtures (Bray 1994) is 
\begin{equation}
\begin{cal} F \end{cal}= \int d {\bf r}\left[ \frac{a}{2}\varphi^2
+\frac{b}{4}\varphi^4+\frac{\kappa}{2} (\nabla \varphi)^{2}\right]
\label{fren}
\end{equation}
where $\varphi$ is the order parameter which describes the normalized 
difference in the 
densities of the two fluids. 
The polynomial terms are related to the bulk
properties of the fluid.  
While the parameter $b$ is always positive,
the sign of $a$ distinguishes between a disordered 
and a segregated mixture.
The case with $a > 0$ gives a polynomial with one minimum in the origin
corresponding to the a state with $\varphi=0$ everywhere
while two minima are present when $a < 0$ corresponding to the two pure phases
with $\varphi = \pm \sqrt{-a/b}$. In the following we
 will consider a deep quench 
(well below the critical value $a=0$) into the coexistence region 
with
$a=-1$ and $b=1$.
The gradient term is related to the interfacial properties.
The  equilibrium profile between the two coexisting bulk phases
is $\displaystyle \varphi(x)=\tanh \sqrt{\frac{1}{2 \kappa}}x$ giving a
 surface tension equal to   $\displaystyle \frac{2}{3} \sqrt{2 \kappa}$
and  an interfacial width
 proportional to $\sqrt{2 \kappa}$ (Rowlinson \& Widom 1982). 
The total fluid density $n$ is not explicitly 
included in (\ref{fren}) because we
are dealing with an incompressible mixture with  $n$ constant
throughout the volume of the fluid and throughout its motion. 

The thermodynamic properties of the fluid 
follow from the free energy (\ref{fren})
(Reichl 1980).
The chemical potential difference between the two fluids is given by
\begin{equation}
\Delta \mu =  \frac{\delta{\cal F}}{\delta\varphi}=
a \varphi + b \varphi^3 - \kappa \nabla^2 \varphi .
\label{mu}
\end{equation}   
The pressure is not a scalar but a tensor $P_{\alpha\beta}$ since interfaces in the
fluid can exert non-isotropic forces (Yang {\it et al.} 1976). The
 scalar part $p_o$ follows directly from (\ref{fren})
\begin{eqnarray}
p_o&=& 
\varphi \; \frac{\delta{\cal F}}{\delta\varphi} - f(\varphi)\nonumber \\
&=&\frac{a}{2}\varphi^2 +\frac{3 b}{4}
\varphi^4 -\kappa \varphi \left( \nabla^2 \varphi \right) -\frac{\kappa}{2}
\left( \nabla \varphi \right)^2
\end{eqnarray}
where $f(\varphi)$ is the free-energy density.
In order to calculate the pressure tensor $P_{\alpha\beta}$, one has
to ensure that $P_{\alpha\beta}$ obeys the condition of mechanical equilibrium
$\partial_{\alpha} P_{\alpha\beta}=0$ (Evans 1979). A suitable choice is 
\begin{equation}
P_{\alpha\beta}=p_o \delta_{\alpha\beta}
+ \kappa \partial_{\alpha} \varphi \partial_{\beta} \varphi.
\label{pres}
\end{equation}

{\it Lattice Boltzmann scheme.} 
We use a square lattice in which each site has eight nearest neighbours. 
The lattice
has horizontal and vertical links of unit length and diagonal links of length
$\sqrt{2}$. 
We  denote by $L$ the number of lattice sites in each direction.  
The variables of the lattice Boltzmann algorithm  are two sets of
distribution functions $f_{i}({\bf r}, t)$ and $g_{i}({\bf r}, t)$,
defined on each lattice site ${\bf r}$ at time $t$.
Each of them is associated with a velocity vector ${\bf e}_i$.
Defined $\Delta t$ as the simulation time step, the quantities
${\bf e}_{i} \Delta t$ are constrained to be  lattice vectors:
$|{\bf e}_{i}| \Delta t=1$ for the horizontal and vertical directions ($i=1,2,3,4$)
and $|{\bf e}_{i}| \Delta t=\sqrt{2}$ for the diagonal directions ($i=5,6,7,8$).
In figure 1 we show a plaquette of the lattice with 
the velocity vectors.
Two functions $f_{0}({\bf r}, t)$ and 
$g_{0}({\bf r}, t)$, corresponding to the distribution components that do not
propagate (${\bf e}_{0}={\bf 0}$), are also taken into account.

The distribution functions are related to the total density  {\it n},  
to the fluid velocity $\bf u$ and to the density difference $\varphi$ through
\begin{equation}   
n=\sum_{i}f_{i} , \hspace{1.3cm} n{\bf u}=\sum_{i}f_{i}{\bf e}_{i} ,\hspace{1.3cm}   
\varphi=\sum_{i}g_{i} .
\label{phys}   
\end{equation}  
They evolve during the time step $\Delta t$ according to   
a single relaxation-time Boltzmann equation (Bhatnagar {\it et al.} 1954,
Chen {\it et al.} 1992):
\begin{eqnarray}   
f_{i}({\bf r}+{\bf e}_{i}\Delta t, t+\Delta t)-f_{i}({\bf r}, t)&=&   
-\frac{1}{\tau}[f_{i}({\bf r}, t)-f_{i}^{eq}({\bf r}, t)], \label{dist1}\\   
g_{i}({\bf r}+{\bf e}_{i}\Delta t, t+\Delta t)-g_{i}({\bf r}, t)&=&   
-\frac{1}{\tau_{\varphi}}[g_{i}({\bf r}, t)-g_{i}^{eq}({\bf r},   
t)], \label{dist2} 
\end{eqnarray}   
where $\tau$ and  $\tau_{\varphi}$ are independent   
relaxation parameters, $f_{i}^{eq}({\bf r}, t)$ and 
$g_{i}^{eq}({\bf r}, t)$ are local equilibrium distribution functions.

The time evolution occurs in two steps: a collision
 and a propagation step. 
Firstly, the distribution functions 
arriving at the same time on the same site change according to
the equation (we will write it only for one of the two distributions)
\begin{equation}   
f_{i}^{c}({\bf r}, t)=f_{i}({\bf r}, t)   
-\frac{1}{\tau}[f_{i}({\bf r}, t)-f_{i}^{eq}({\bf r}, t)].
\label{coll}
\end{equation}
Then the distribution functions are moved along the lattice directions
according to the rule
\begin{equation}
f_{i}({\bf r}+{\bf e}_{i}\Delta t, t+\Delta t)=f_{i}^{c}({\bf r}, t).
\label{prop}
\end{equation}
Equation (\ref{dist1}) comes from   
combining 
equations (\ref{coll}) and (\ref{prop}).

The system is initialized with $n=1$, $\varphi=0$ and 
${\bf u}={\bf 0}$ everywhere. The choice of a constant
value of $n$  is justified  by the incompressibility
of  the mixture. Our simulations  
confirm that $n$ stays constant throughout the system during all the 
time evolution. 
The initial value of $\varphi$ 
corresponds to  a symmetric 50:50 mixture
 with the two fluids completely mixed.
These quantities are locally conserved in any collision process and, 
therefore, we
 require that the local equilibrium distribution functions fulfil the equations 
\begin{eqnarray}
\sum_i (f_i^{eq}-f_i)=0 &\Rightarrow & \sum_i f_i^{eq}=n \nonumber \\
\sum_i (g_i^{eq}-g_i)=0 &\Rightarrow&\sum_i g_i^{eq}=\varphi \label{req1} \\
\sum_i (f_i^{eq}-f_i){\bf e}_i={\bf 0} &\Rightarrow&\sum_i f_i^{eq} {\bf e}_i=
n {\bf u}\nonumber
\end{eqnarray}
Following Orlandini {\it et al.} (1995) and Swift {\it et al.} (1996) 
the higher moments of the local 
equilibrium distribution functions are defined so that we 
can obtain continuum equations pertinent to a binary fluid mixture.  
We define  
\begin{equation}   
\sum_{i}f_{i}^{eq}e_{i\alpha}e_{i\beta}=P_{\alpha\beta}+n u_{\alpha} u_{\beta} \;,   
\label{eqn0}
\end{equation}  
\begin{equation}  
 \sum_{i}g_{i}^{eq}e_{i\alpha}=\varphi u_{\alpha} \; ,
\label{eqn} 
\end{equation}
\begin{equation}
\sum_{i}g_{i}^{eq}e_{i\alpha}   
e_{i\beta}=\Gamma \Delta\mu\delta_{\alpha\beta}+\varphi   
u_{\alpha}u_{\beta} \;.
\label{eqn6}  
\end{equation}   
where $\Gamma$ is a coefficient related to the mobility of the fluid.
We are considering a mixture with the two fluids having the same mechanical
properties. 
The constraint (\ref{eqn}) expresses the fact that  the two 
fluids  have the same velocity. 
The local 
equilibrium distribution functions can be expressed as 
 an expansion in terms of the 
velocity ${\bf u}$ (Orlandini {\it et al.} 1995, Swift {\it et al.} 1996):
\begin{eqnarray}
f_0^{eq}&=& A_0+C_0 u^2 \nonumber\\
f_i^{eq}&=& A_I+B_I u_\alpha e_{i\alpha}+C_I u^2+D_I u_\alpha u_\beta
e_{i\alpha} e_{i\beta}+ G_{I,\alpha\beta}e_{i\alpha} e_{i\beta}
\;\;\;\; i=1,2,3,4 \label{svil1}\\
f_i^{eq}&=& A_{II}+B_{II} u_\alpha e_{i\alpha}+C_{II} u^2
+D_{II} u_\alpha u_\beta
e_{i\alpha} e_{i\beta}+ G_{II,\alpha\beta}e_{i\alpha} e_{i\beta}
\;\;\;\; i=5,6,7,8 \nonumber
\end{eqnarray}
Similarly
\begin{eqnarray}
g_0^{eq}&=& H_0+J_0 u^2 \nonumber\\
g_i^{eq}&=& H_I+K_I u_\alpha e_{i\alpha}+J_I u^2+Q_I u_\alpha u_\beta
e_{i\alpha} e_{i\beta}
\;\;\;\; i=1,2,3,4 \label{svil2}\\
g_i^{eq}&=& H_{II}+K_{II} u_\alpha e_{i\alpha}+J_{II} u^2
+Q_{II} u_\alpha u_\beta
e_{i\alpha} e_{i\beta}
\;\;\;\; i=5,6,7,8 \nonumber
\end{eqnarray}
The relations (\ref{req1})-(\ref{eqn6})
can be used to fix the coefficients of these expansions.
The results are reported in the Appendix.

{\it Continuum equations.}
It has been shown by Orlandini {\it et al.} (1995) and Swift {\it et al.} (1996)
that the above 
described lattice Boltzmann scheme simulates the continuity, 
the incompressible
Navier-Stokes and the convection-diffusion equations
\begin{eqnarray}
\partial_{\alpha} u_{\alpha} &=& 0 \label{concon}\;\;,\\
\partial_{t} u_{\alpha}+u_{\beta}\partial_{\beta}    
u_{\alpha}&=& -\frac{1}{n}\partial_{\alpha}p_{o}+\nu \nabla^{2} u_{\alpha} \;\;, \\   
\partial_{t}\varphi+\partial_{\alpha}(\varphi   
u_{\alpha})&=&\Gamma \Theta \nabla^{2} \frac{\delta{\cal F}}{\delta\varphi}\;\;.   
\label{conv1}
\end{eqnarray}   
A Chapman-Enskog expansion (Chapman \& Cowling 1970, Swift {\it et al.} 1996)
of equations (\ref{dist1}) and (\ref{dist2}) to $O (\Delta t^{2})$ allows 
to show that 
\footnote 
{Actually, spurious terms appear in equations (\ref{concon})-(\ref{conv1}),
whose effects are shown negligible in the papers of Orlandini {\it et al.} (1995)
and Swift {\it et al.} (1996).}
\begin{equation}   
\nu=\frac{(2\tau-1)}{6}(\Delta t),\qquad \Theta=\Delta t(\tau_{\varphi}-\frac{1}{2}).
\label{param}   
\end{equation}   
We choose $\tau_{\varphi}=(1+1/\sqrt{3})/2$   
in order to minimize the correction terms of order $\Delta t^{3}$.
Thus we are left with two free variables
$\tau$ and $\Gamma$ which control the kinematic
 viscosity $\nu$ and the macroscopic
mobility $\Gamma \Theta$, respectively. 
All the simulations were run with $\kappa=1$
(this choice corresponds to an interfacial width of approximately three lattice
spacings), $\Gamma \Theta=0.2$ and units in which $\Delta t=1$.  

\section{Shear boundary conditions}
\par
The next step in implementing the algorithm is to apply a shear flow on the
system. 
We have modified previous lattice Boltzmann schemes
for a single fluid where
sliding walls moving in a lattice direction were used to  enforce
 the fluid a given velocity. 
The flow is assumed to be directed along the $x$-axis (horizontal
 on the lattice). In this direction we considered
 periodic boundary conditions. Then
departing distribution functions on outward-pointing links 
re-enter the lattice
via corresponding inward-pointing links on the opposite boundary
 with no constraint
on the macroscopic velocity or density.
The walls are on the top and the bottom of the lattice.
 Here  one has to
face two problems: 
the correct value of the velocity has to be
imposed to the fluid, and the distribution functions pointing outward have
to be managed in such a way that mass conservation 
for both components and momentum 
conservation for the bulk are always guaranteed. 
In the following we  discuss
two possible ways for introducing the  sliding walls \footnote
{A different approach based on 
 Lees-Edwards boundary conditions (Lees \& Edwards (1972)) has been used
by  Wagner \& Yeomans (1999). 
These boundary conditions, widely used in molecular dynamics, 
 identify the point at $(x,y)$ with the point at 
$(x+\gamma L \Delta t,y+L)$, where $\gamma$ is the shear rate. 
Lattice Boltzmann implementation of Lees-Edwards boundary
conditions requires 
also to change the usual  definition of the local equilibrium
distribution functions so that  
we have preferred to use the more straightforward
approach with  the sliding walls.}. 

{\it Equilibrium scheme.}
In the first scheme that we have considered, the
boundary walls are placed on the links 
beyond the upper and the lower 
sites at a distance equal to half
lattice spacing, in the spirit of some works on shear boundary conditions in
lattice
Boltzmann methods for one fluid  (Cornubert {\it et al.} 1991, Ladd 1994).
We assign to the local equilibrium distributions at the top (t) 
and the bottom (b)
of the lattice the values
corresponding to the velocities
$\displaystyle w_{x,t}=\gamma \frac{L-1}{2}$, 
$w_{y,t}=0$, 
$\displaystyle w_{x,b}=-\gamma \frac{L-1}{2}$, 
$w_{y,b}=0$,
being $L-1$ the distance between the top and the bottom
of the lattice (He {\it et al.} 1997). This method is 
expected to be accurate
for relaxation parameters near unity, since when  
$\tau=1$ the collision process
(see equation (\ref{coll})) simply replaces the distribution function with the
 local equilibrium
value. For describing the propagation step let us 
refer to the upper row  and 
to the $f_i$'s. 
After a collision, for each site of the upper row,  
there are three distributions $f_5^{c}(t)$, $f_2^{c}(t)$,  
$f_6^{c}(t)$
 pointing outward the system. 
The propagation  is implemented with the following scheme:
\begin{eqnarray}
f_{7}({\bf r}+{\bf e}_{3}\Delta t, t+\Delta t)&=&f_{6}^{c}({\bf r}, t) \nonumber \\
f_{4}({\bf r}, t+\Delta t)&=&f_{2}^{c}({\bf r}, t)  \\
f_{8}({\bf r}+{\bf e}_{1}\Delta t, t+\Delta t)&=&f_{5}^{c}({\bf r}, t) \nonumber
\label{1thbis}
\end{eqnarray}
This choice can be 
 justified considering reflection of particles against 
the boundary wall.
This allows to
guarantee the conservation of  mass and momentum, 
that on each site 9 distribution
functions are still defined, and
that propagation between adjacent sites on the boundary
occurs simultaneously to the propagation on all the inner sites, since
$f_5^{c}(t)$ and $f_6^{c}(t)$ are propagated over a distance $\sqrt 2$
in a time step and $f_2^{c}(t)$  over a distance equal to one.
This method, however, 
as  expected and also shown in the following, does not work well
 for arbitrary values of the
relaxation parameters. Therefore we heve  developed a second way
 for introducing the sliding walls.

{\it Collisional scheme.} This second method is an improvement of 
a  scheme proposed by Zou \& He (1997) for one-fluid systems. 
In this case
the boundary walls are placed on the upper and lower rows of sites 
 (see, e.g., Noble {\it et al.} 1995, Inamuro {\it et al.} 1995).
We start from the propagation step, as it is realized in the original 
version by  Zou \& He (1997),
 and consider again the sites at the top 
of the lattice.
After the propagation the 
functions $f_0(t)$, $f_1(t)$, $f_5(t)$, $f_2(t)$, $f_6(t)$ and $f_3(t)$ are known on
each site of the upper row.
One uses equations (\ref{phys}) to determine $f_7(t)$, $f_4(t)$, $f_8(t)$ and $n$.
Requiring that the wall velocities 
$\displaystyle w_{x,t}=\gamma \frac{L-1}{2}$,
$w_{y,t}=0$
are imposed to the fluid, we can write
\begin{eqnarray}
f_7(t)+f_4(t)+f_8(t)&=&n - \left [f_0(t)+f_1(t)+f_5(t)+f_2(t)+f_6(t)+f_3(t)
\right ] \nonumber \\
f_8(t)-f_7(t)&=&n \;\gamma \;\frac{L-1}{2} - \left [f_1(t)-f_3(t)+f_5(t)-f_6(t)
\right ] \label{star}\\
f_7(t)+f_4(t)+f_8(t)&=&f_5(t)+f_2(t)+f_6(t)\nonumber
\end{eqnarray} 
Consistency of equations (\ref{star}) gives
\begin{equation}
n = f_0(t)+f_1(t)+f_3(t)+ 2 \left [f_2(t)+f_6(t)+f_5(t)\right ]
\label{cons}
\end{equation}
The system of equations (\ref{star}) reduces to two equations with three unknown
variables.
To close the system of equations the bounce-back rule (Lavall\'{e}e {\it et al.} 1991,
Cornubert {\it et al.} 1991) is adopted for  the 
distribution functions
 normal to the boundary. This means that the value of $f_4(t)$ is fixed
assigning to it the known value of its outward-pointing counterpart $f_2(t)$. 
Then  one can solve the system of equations (\ref{star})
finding the solutions
\begin{eqnarray}
f_4({\bf r},t)&=&f_2({\bf r},t) \nonumber \\
f_8({\bf r},t)&=&f_6({\bf r},t)-\frac{1}{2}\left [ f_1({\bf r},t)-f_3({\bf r},t)
\right ]
+ \frac{1}{2} \; n \;\gamma \;\frac{L-1}{2} \label{phi} \\
f_7({\bf r},t)&=&f_5({\bf r},t)+\frac{1}{2}\left [ f_1({\bf r},t)-f_3({\bf r},t)
\right ]
- \frac{1}{2} \; n \;\gamma \;\frac{L-1}{2} \nonumber
\end{eqnarray} 
With this choice for the inward-pointing distributions the desidered momentum at the
boundary is achieved. At this point the collision step is applied to 
all sites, 
including the boundary ones. Unfortunately this scheme does not allow 
to conserve 
exactly the total  mass due to the fact that the distribution 
functions $f_5^{c}(t)$,
$f_2^{c}(t)$ and $f_6^{c}(t)$ resulting from the collision step
are not accounted for any more, and this avoids the exact mass conservation 
(Zou \& He 1997).

We have improved the above scheme in order to overcome the mass
conservation problem. We consider again the propagation step and the problem of
calculating the unknown quantities $f_7(t)$,
$f_4(t)$ and $f_8(t)$.
After the propagation
step on each upper site one has the distribution functions 
$f_1(t)$, $f_5(t)$, $f_2(t)$, $f_6(t)$ and $f_3(t)$ 
coming from the nearest sites,
$f_0(t-\Delta t)$, which does not propagate, and 
$f_5(t-\Delta t)$, $f_2(t-\Delta t)$ and $f_6(t-\Delta t)$, all 
referred to the previous time step, which 
 were lost in the version of Zou \& He (1997).
Mass will be conserved if the total density 
$n$ on each site is equal to the quantity $\hat{n}$
given by the sum
\begin{eqnarray}
\hat{n}(t,t-\Delta t)\:\:&=&\:\: f_0(t-\Delta t)+f_5(t-\Delta t)+f_2(t-\Delta t)+ 
f_6(t-\Delta t) \nonumber\\
&&+f_1(t)+f_5(t)+f_2(t)+f_6(t)+f_3(t) .
\label{hat}
\end{eqnarray}
We also require that equations (\ref{star}) are fulfilled. 
In order to impose
the constraint that on all the boundary sites $n=\hat n$, we have to introduce
an extra degree of freedom in the system of equations. We have choosed $f_0(t)$
since it does not propagate.
The solutions of the system of equations (\ref{star}),  $n=\hat n$
 and (\ref{hat}) are
\begin{equation}
f_0({\bf r},t)=\hat{n}- \left [f_1({\bf r},t)+f_3({\bf r},t) \right ]
-2 \left [ f_2({\bf r},t)+
f_5({\bf r},t)+f_6({\bf r},t)\right ]
\label{f0}
\end{equation}
and again  the (\ref{phi}).
In this way it is guaranteed that
\begin{eqnarray}
n({\bf r},t)&=&\hat{n}({\bf r},t,t-\Delta t)\nonumber \\
n u_x({\bf r},t)&=&n \;\gamma \;\frac{L-1}{2}\\
n u_y({\bf r},t)&=&0 \qquad . \nonumber
\end{eqnarray} 
By this procedure, once the system has been initialized, the application of 
the propagation and collision steps goes on 
preserving mass and momentum conservation and implementing
the correct velocity values on the boundaries, as it has also been
verified numerically.

{\it Comparison between the two schemes}. 
Here we compare  the validity of the two schemes. 
We have
studied how the steady state  velocity profile 
is   reached and we have measured the slip velocity. 
The shear is applied as usual from the beginning of the phase 
separation.
Preliminarly we have checked that the boundary conditions
do not introduce any artificial 
 discontinuity on the walls. Therefore we
have considered 
the behaviour of the densities $n$ and $\varphi$ along vertical sections
normal to the flow direction. 
In figure 2 (a)-(b) the plots of $n$ and $\varphi$,
for the  middle section of the system, 
are showed for the collisional scheme. 
It can be seen that the total density
$n$ stays constant and equal to 1 all over the system
 (its deviations from 1 are less than $1/10^{5}$)
and no boundary effects can be
observed.
 The profile of $\varphi$, with the fluctuations corresponding 
to the presence of interfaces,
 does not show anything pathological, too.
Similar results have been obtained 
with the equilibrium scheme.
In figure 3 
we show 
at consecutive times the $x$-component of the fluid
velocity along a vertical line in the middle of the system
calculated with the collisional scheme.
The velocity profiles are independent on the
particular vertical section considered.
The steady velocity profile is the planar
Couette flow (Landau \& Lifshitz 1959)
with the $x$-component
of the velocity having a  linear dependence on the $y$ coordinate. 
The evolution of the velocity profile in  figure 3 is 
very similar to that observed
in simple fluids (Schlichting 1979, p. 92).
The same evolution is obtained with the equilibrium scheme when $\tau = 1$.
The situation is different for larger values of $\tau$,
when, with the first scheme,   the time needed to reach the steady 
velocity profile is much longer. Actually, it happens that for 
  large enough values of 
$\tau$ the linear profile is not reached before finite-size effects become relevant in the simulations.
 This can be understood from equation (\ref{coll}). Indeed, 
if $\tau > 1$,
$f_i^{c}({\bf r},t)$ takes more time to reach the local equilibrium value 
$f_i^{eq}({\bf r},t)$ which contains the information about 
the velocity of walls. 

A measure of  
the time $t_r$ required to reach the steady state velocity profile
can be done in this way.
We say that the steady state is reached if 
\begin{equation}
\frac{\sum_{{\bf r}}|u_x({\bf r},t+\Delta t)-u_x({\bf r},t)|+
|u_y({\bf r},t+\Delta t)-u_y({\bf r},t)|}
{\sum_{{\bf r}}|u_x({\bf r},t)|+|u_y({\bf r},t)|}<T
\end{equation}
where $T$ is a tolerance set to $10^{-6}$. 
In figure 4 we show the results obtained with the collisional scheme.
The time $t_r$ required to reach the steady state
decreases at increasing values of $\tau$. 
The analytic solution of the Navier-Stokes equation for a single fluid
subject to a  Couette flow suggests a similar behaviour.
For a single fluid it can be shown that 
the non-stationary part of the 
velocity profile behaves as a series of exponential terms
 $\exp({-\pi^2 m^2 \nu t /L^2)} $ with $m$ an integer
  (Schlichting 1979).
For this reason   we have reported our results 
 in a log-log scale,
even if in this plot we do not observe a linear behaviour.

Next we consider the behaviour
of the slip velocity. It is   defined as the difference between the wall 
velocity and the fluid velocity along the
wall itself and  should be negligible. The slip velocity can be 
measured by 
the maximum relative error.
At the top boundary it can be calculated as
\begin{equation}
E_m=max|\frac{u_x(i,L)-w_{x,t}}{w_{x,t}}| \;\;\;\;\; i=1,2,...,L
\end{equation}
where again $u_x(i,L)$ is the fluid velocity at the $i$-th site along the $x$ 
direction at
the top boundary and $w_{x,t}=\gamma (L-1)/2$ is the wall velocity. 
The results of this measure for different
values of $\tau$ and for both  schemes  are 
presented in figure 5.
For the collisional scheme $E_m \simeq 10^{-5}$ over the explored
range $[0.7, 20]$ of values of $\tau$ (figure 5 (a)). 
In the equilibrium scheme 
$E_m$ has a minimum value at $\tau=1$ and then
increases with  $\tau$  (figure 5 (b))
arriving to  values of order 0.3.
The conclusion is that 
the first scheme gives less convincing results
and we have preferred to use
the collisional  scheme to run our simulations on phase separation.

\section{Simulations of phase separation}
\par
We have studied the dynamics of phase separation 
on systems with size $L=256$, using 
the values written  in section 2
 for the parameters of the model and varying 
 $\gamma$ in the range $[0.001, 0.01]$  and $\tau$ in
 $[0.7, 5]$, 
being mainly interested in the hydrodynamic regime at low 
viscosity.

Here we report the results obtained for $\tau=0.7$, which corresponds 
to a viscosity $\nu=0.067$, and $\gamma=0.005$. Similar results 
have been obtained for  other values of parameters.
A sequence of configurations at different values of
the shear strain $\gamma t$ is shown in the left column of figure 6. 
The first step of the time evolution is the formation of well defined 
interfaces. Then, for a while,  domains grow isotropically
as in the case without shear. At $\gamma t=0.4$, for example,
the effects of the shear can be observed 
  only near the moving walls due to the fact that at this time
the linear profile of the velocity is not yet reached and the velocity
is greater close to the walls (see figure 3).
The deformation induced by the flow
can be observed everywhere in the system starting from values  
$\gamma t \simeq 1$ . Later on  the domains 
become very stretched
 and progressively aligned with the flow direction as it can be clearly
seen  at $\gamma t = 7$ and $11$. It is also clear,
by looking at these two configurations, that the system shows inhomogeneities
and in particular domains with different thicknesses. 

A quantitative analysis of the length scales present in the system
can be done  by considering the behaviour of the 
structure factor $C({\bf k},t)$.
This quantity is the one of experimental interest being accessible 
through scattering techniques. 
It corresponds to  the Fourier transform of the two-points
correlation function and can be  computed numerically as $<\varphi({\bf k},t) 
\varphi({-\bf k},t)>$, where $\varphi({\bf k},t)$ is the Fourier transform of the 
order parameter
 $\varphi({\bf r},t)$ and $<...>$ refers to an average over 
an ensemble of different initial
realizations of the system. 
The wave vector ${\bf k}$ in the reciprocal lattice is given by 
$\displaystyle {\bf k}=\frac{2 \pi}{L}(n_x {\bf i}+n_y {\bf j})$, where ${\bf i}$ 
and ${\bf j}$ are two unit vectors in the $k_x$ and $k_y$ directions, respectively, and
$n_x$ and $n_y$ range from $-L/2$ to $L/2-1$.
It could be useful to remind the behaviour of  $C({\bf k},t)$ 
in the case without shear. Its shape is that of a volcano with a contour plot
similar to that shown in figure 6 at $\gamma t =0.4$. 
The radius of the volcano 
defines the inverse of a  characteristic length representing
 the average size of domains at a given time.
As time goes by,
the height of this volcano increases while its support shrinks 
towards the origin. This means that 
 the characteristic length scale associated with the 
 size of domains increases and the 
 system  is more and more ordered on that scale.
The fact that the structure factor
is circular reflects the isotropy of the system. 

The  shear affects  in a significative way the evolution 
of the structure factor. 
At $\gamma t =1$  the shape of the volcano is deformed into an 
 elliptical structure as it can be seen in the contour plot of figure 6.
Moreover also the profile of the edge of the volcano  is 
deformed and two maxima can  be observed distributed 
around  the main diagonal
of  the $k_x-k_y$ plane.
In the further evolution the two broad maxima at 
 $\gamma t =1$ become divided  into two separate foils and on each foil 
two different peaks can be observed. This can be clearly seen 
at $\gamma t = 7 $.
These two foils becomes closer and closer to the axis $k_x = 0$
and increasingly more aligned along the $k_y$ axis.
This pattern corresponds to interfaces very elongated in the direction 
of the flow  as it appears
 in the corresponding
pictures of configurations.

The presence of two peaks on each foil  implies  the existence of 
different length scales in the systems.
Due to the symmetry
$C({\bf k},t)=C({-\bf k},t)$ the two foils are perfectly symmetric
and it is sufficient to consider only the two peaks of one foil.
Therefore for  each direction there are two relevant
 length scales.
Focussing on what happens in the direction normal to the flow  
we arrive to the conclusion that there are domains with two 
 characteristic thicknesses in the system. This better explains
the inhomogeneities observed in the configurations at $\gamma t = 7$ and $11$.

Also the dynamical evolution of the  4-peaked
 structure factor is very peculiar.
At $\gamma t=7$ the peaks which  prevail are the ones
 with a larger value of $k_y$. This corresponds to a prevalence
of thin domains in the system, as it can perhaps be observed by
directly looking at the corresponding configuration. 
Later on the more isotropic peak with the smaller value
of $k_y$,
which has grown faster than the other, becomes higher.
This can be observed in figure 6 at  $\gamma t=11$
in correspondence of a  larger abundance 
of thicker domains.

The physical  picture of what happens in the system is the following.
By stretching the interfaces the shear  induces a stress in the system.
This stress increases more and more producing  a prevalence of thin domains.
 At a certain point  
the domains are 
broken by the flow in points where the stress is larger.
 This occurs in a cooperative way with a 
sequence of ruptures producing domains which are less
stretched.
Then the thicker domains, not yet broken by the flow, become more
numerous and the more isotropic peak of $C({\bf k},t)$
 dominates as at $\gamma t=11$. 
It was not possible to follow longer in our simulations 
the alternate predominance of the two peaks due to inevitable
appearance of finite-size effects.
Previous analytical results of  Corberi {\it et al.} (1998) show 
that in the diffusive regime
 the alternate prevalence of the two  peaks 
  continues periodically on a logarithmic time scale. 
We cannot give results on this.
Here we mention that in experiments with polymer solutions by
Migler {\it et al.} (1996), 
after a shallow quench under the critical temperature,
segregation is observed 
with 4-peaked structure factors  similar to those of figure 6.
Moreover, in previous experimental papers by Laufer {\it et al.}
(1973) and Mani {\it et al.} (1991)-(1992)
always on polymer solutions, the description of the dynamics 
of the network of domains is similar to that given above.
We believe that our simulations, where for the first time the
presence of two length scales is considered,
 help to understand these experimental facts.

We conclude this section giving the results of an explicit 
evaluation of the size of the domains.  This measure is 
usually obtained from the momenta of the 
structure factor. Since our system is anisotropic 
we define the  average size of domains in the
flow direction as 
\begin{equation}
R_x(t) = \pi \frac{\int d{\bf k} C({\bf k},t)}{\int d{\bf  k} |k_x| C({\bf k},t)} 
\end{equation}
and analogously for $R_y(t)$, in the shear direction normal to the flow. 
They are one half of the wave length corresponding
to the characteristic wave vector.
In the case without shear
lattice Boltzmann simulations with  our
parameters  give $R_x = R_ y \sim t^{2/3}$ 
(Osborn {\it et al.} 1995).  $\alpha=2/3$ is the value of the growth exponent
typical of the inertial regime which occurs at very low viscosity 
(Furukawa 1985).
From our simulations we extract the behaviour shown in figure 7.
As in all the other
simulations with a shear flow (Yeomans 1999), 
it has not been  possible to extract
the value of the exponent $\alpha$ due to the relevance 
of finite-size effects 
\footnote {An analytical determination of 
the exponent  $\alpha$ is only available for the diffusive regime
 (Corberi {\it et al.} (1998) , Rapapa \& Bray (1999)).
 It is found that  $\alpha=4/3$ in the flow direction,
 and $\alpha=1/3$ in the other
directions
as in the case without shear.}.
Using an argument proposed by Corberi {\it et al.} (1999), based on a renormalization
group approach (Bray 1990),
one should expect in the present case, if dynamical scaling for the structure
factor holds, that 
the growth in the shear direction obeys the same
power law as in the case without shear, $R_y(t) \sim t^{2/3}$, 
while in the $x$-direction 
$R_x(t) \sim \gamma t R_y(t) \sim \gamma t^{5/3}$. 
This means that the growth exponent $\alpha_x$ in the flow direction is related
to $\alpha_y$, the growth exponent in the shear direction,  by the relation
$\alpha_x=\alpha_y+1$, where the extra contribution $1$ with respect to the unsheared
case comes from the convective term in the convection-diffusion equation (\ref{conv1}).
Our runs, performed on an Alpha Workstation XP1000 with 1 GB of RAM, 
do not allow to check these predictions.
The results of figure  7 can be more easily related to the above discussed
dynamical behaviour  of the structure factor
with the alternate dominance of  the two peaks  
which corresponds to   the oscillatory pattern  of 
$R_x$ and $R_y$.   
We also observe that these
oscillations make more difficult a possible evaluation 
of the growth exponents.
\section{Conclusions}
\par
We studied phase separation in sheared binary fluid mixtures 
at low viscosity. 
We introduced proper boundary conditions to impose the shear flow on the
system. These boundary conditions strictly conserve mass and momentum
and do not introduce any appreciable slip velocity.
Our main results show that the network of domains
is characterized by the existence  of two typical length scales
for each direction. Indeed, domains with different thickness
 are clearly
visible in the simulations.  
Structure factors with four peaks confirm these observations. 
During the time evolution there is 
 an alternate  dominance of two of these   peaks over the other couple.
This corresponds to a larger abundance first of thin 
and later of thicker domains.
Simulations on much larger scales are needed to establish 
if the oscillations of the peaks  only characterize
 an initial transient or if they continue indefinitely,
as it is suggested by the results of Corberi {\it et al.} (1998) 
valid  for the diffusive
regime at infinite viscosity.
 
\section{Ackowledgements}
\par
We thank A. Bonfiglioli for useful preliminary discussions.
G.G. acknowledges support by MURST (PRIN97).
\newpage
\noindent
{\Large {\bf Appendix}}
\vspace{1.0cm}
\par\noindent
A suitable choice of the coefficients in the expansions 
(\ref{svil1})-(\ref{svil2})
consistent with the conditions (\ref{req1})-(\ref{eqn6}) is
\begin{equation}
A_0=n-20 A_{II}, \hspace{0.5cm} A_I=4 A_{II}, \hspace{0.5cm} 
A_{II}=\frac{P_{xx}+P_{yy}}{24}
\label{a's}
\end{equation}
\begin{equation}
B_I=4 B_{II}, \hspace{0.5cm} 
B_{II}=\frac{n}{12}
\label{b's}
\end{equation}
\begin{equation}
C_0=-\frac{2 n}{3}, \hspace{0.5cm} C_I=4 C_{II}, \hspace{0.5cm} 
C_{II}=-\frac{n}{24}
\label{c's}
\end{equation}
\begin{equation}
D_I=4 D_{II}, \hspace{0.5cm} 
D_{II}=\frac{n}{8}
\label{d's}
\end{equation}
\begin{equation}
\hspace{-2cm} G_{I,\alpha\beta}\!=\!4 G_{II,\alpha\beta}, \hspace{0.2cm}
 G_{II,xx}\!=\!-G_{II,yy}\!=\!\frac{P_{xx}\!-\!P_{yy}}{16}, \hspace{0.2cm}
G_{II,xy}\!=\!-G_{II,yx}\!=\!\frac{P_{xy}}{8}
\label{g's}
\end{equation}
\begin{equation}
H_0=\varphi-20 H_{II}, \hspace{0.5cm} H_I=4 H_{II}, \hspace{0.5cm} 
H_{II}=\frac{\Gamma \Delta\mu}{12} 
\label{h's}
\end{equation}
\begin{equation}
K_I=4 K_{II}, \hspace{0.5cm} 
K_{II}=\frac{\varphi}{12}
\label{k's}
\end{equation}
\begin{equation}
J_0=-\frac{2 \varphi}{3}, \hspace{0.5cm} J_I=4 J_{II}, \hspace{0.5cm} 
J_{II}=-\frac{\varphi}{24}
\label{j's}
\end{equation}
\begin{equation}
Q_I=4 Q_{II}, \hspace{0.5cm} 
Q_{II}=\frac{\varphi}{8}
\label{q's}
\end{equation}
\newpage
\noindent
{\Large {\bf References}}
\vspace{1.0cm}
\par\noindent
Bhatnagar, P., Gross, E. P. \& Krook, M. K. 1954 Phys. Rev.
{\bf 94}, 511.\\
Bray, A. J. 1990 Phys. Rev. B {\bf 41}, 6724.\\
Bray, A. J. 1994 Adv. Phys. {\bf 43}, 357.\\
Cates, M. E., Kendon, V. M., Bladon, P. \& Desplat, J. C. 1999
Faraday Discussions {\bf 112}, 1.\\
Chapman, S. \& Cowling, T. 1970 {\sl The Mathematical Theory of
Non-uniform Gases} (Cambridge University Press, Cambridge).\\
Chen, H., Chen, S. \& Matthaeus, W. 1992 Phys. Rev. A {\bf 45},
R5339.\\
Chen, S. \& Doolen, G. D. 1998 Annu. Rev. Fluid Mech. {\bf 30}, 329.\\
Corberi, F., Gonnella, G. \& Lamura, A. 1998 Phys. Rev. Lett. 
{\bf 81}, 3852.\\
Corberi, F., Gonnella, G. \& Lamura, A. 1999 Phys. Rev. Lett. {\bf 83}, 4057.\\
Cornubert, R., d'Humieres, D. \& Levermore, D. 1991 Physica D {\bf 47}, 241.\\
Evans, R. 1979 Adv. Phys. {\bf 28}, 143.\\
Furukawa, H. 1985 Adv. Phys. {\bf 34}, 703.\\
Gonnella, G., Orlandini, E. \& Yeomans, J. M. 1997 Phys. Rev. Lett. {\bf 78}, 1695.\\
Gunton, J. D., San Miguel, M. \& Sahni, P. S. 1983 in 
{\sl Phase Transitions and Critical Phenomena}, edited by C. Domb, J. L. Lebowitz
(Academic, New York).\\
Hashimoto, T., Matsuzaka, K., Moses, E. \& Onuki, A. 1995 Phys. Rev. Lett. {\bf 74}, 
126.\\
He, X., Zou, Q., Luo, L. S. \& Dembo, M. 1997 J. Stat. Phys. {\bf 87}, 115.\\
Inamuro, T., Yoshino, M. \& Ogino, F. 1995 Phys. Fluids {\bf 7}, 2928.\\
Kendon, V. M., Desplat, J. C., Bladon, P. \& Cates, M. E. 1999  
Phys. Rev. Lett. {\bf 83}, 576.\\
Ladd, A. J. 1994 J. Fluid Mech. {\bf 271}, 285.\\
Landau, L. D. \& Lifshitz, E. M. 1959 {\sl Fluid Mechanics} (Pergamon Press, Oxford).\\
Laufer, Z., Jalink, H. L. \& Staverman, A. J. 1973 Journal of Polymer science
{\bf 11}, 3005.\\
Lavall\'{e}e, P., Boon, J. \& Noullez. A. 1991  Physica D {\bf 47}, 233.\\
Lees, A. W. \& Edwards, S. F. 1972 J. Phys. C  {\bf 5}, 1921.\\
Mani, S., Malone, M. F., Winter, H. H., Halary, J. L. \&
Monnerie, L. 1991 Macromolecules {\bf 24}, 5451.\\
Mani, S., Malone, M. F. \& Winter, H. H.
1992 Macromolecules {\bf 25}, 5671.\\
Migler, K., Liu, C. \& Pine, D. J. 1996 Macromolecules {\bf 29}, 1422.\\
Noble, D. R., Chen, S., Georgiadis, J. G. \& Buckius, R. O. 1995
Phys. Fluids {\bf 7}, 203.\\
Ohta, T., Nozaki, H. \& Doi, M. 1990 Phys. Lett. A {\bf 145}, 304;
J. Chem. Phys. {\bf 93}, 2664.\\
Olson, J. F. \& Rothman, D. H. 1995 J. Stat. Phys. {\bf 81}, 199.\\
Onuki, A. 1997 J. Phys. Condens. Matter {\bf 9}, 6119.\\
Orlandini, E., Swift, M. R. \&  Yeomans, J. M. 1995
Europhys. Lett. {\bf 32}, 463.\\
Osborn, W. R., Orlandini, E., Swift, M. R., Yeomans, J. M. \&
Banavar, J. R. 1995 Phys. Rev. Lett. {\bf 75}, 4031.\\
Padilla, P. \& Toxvaerd, S. 1997 J. Chem. Phys. {\bf 106}, 2342.\\
Rapapa, N. P. \&  Bray, A. J. 1999 Phys. Rev. Lett. {\bf 83}, 3856.\\
Reichl, L. E. 1980 {\sl A Modern Course in Statistical
Physics} (Arnold, London).\\
Rothman, D. H. 1991 Europhys. Lett. {\bf 14}, 337.\\
Rowlinson, J. S. \& Widom, B. 1982 {\sl Molecular Theory of Capillarity}
(Clarendon Press, Oxford).\\
Schlichting, H. 1979 {\sl Boundary Layer Theory}
 (Mc Graw Hill series in Mechanical Engineering).\\
Shou, Z. \& Chakrabarti, A. 2000 Phys. Rev. E {\bf 61}, R2200.\\
Swift, M. R.,  Orlandini, E., Osborn, W. R. \& Yeomans, J. M. 1996
Phys. Rev. E {\bf 54}, 5041.\\
Wagner, A. J. \& Yeomans, J. M. 1999 Phys. Rev. E {\bf 59}, 4366.\\
Wu, Y. N., Skrdla, H., Lookman, T. \& Chen, S. Y. 1997 
Physica A {\bf 239}, 428.\\
Yang, A. J. M., Fleming, P. D. \& Gibbs, J. H. 1976
J. Chem. Phys. {\bf 64}, 3732.\\
Yeomans, J. M. 1999 preprint OUTP-99-06S, Ann. Rev. Comp. Physics, ed. D. Stauffer,
in press.\\
Zou, Q. \& He, X. 1997 Phys. Fluids {\bf 9}, 1591.
\newpage
\noindent\Large\textbf{Figure Captions}
\normalsize
\vspace{1.0cm}
\begin{description}
\item{Figure 1}: 
One plaquette of the lattice used in our model with the velocity
vectors ${\bf e}_i$ ($\Delta t=1$).
\item{Figure 2}: 
Plots of the total fluid density $n$ (a) and of the density difference $\varphi$
(b) along a vertical section 
at $t=32$, using the collisional scheme.
Simulations were run with $L=128$, $\gamma=0.01$ and $\tau=1$.
\item{Figure 3}: 
Plot of the $x$-component of the fluid velocity
as a function of the $y$ coordinate at consecutive times: 
$(\triangle) ~ t=2$, $(\circ) ~ t=8$, $(\star) ~ t=16$, $(\bullet) ~ t=32$
for the collisional sheme of shear boundary conditions.
\item{Figure 4}: The time $t_r$ 
required to reach the steady state  is shown as a function
of the relaxation parameter $\tau$, using the collisional scheme. 
\item{Figure 5}: The maximum relative error $E_m$
 in the velocities at the top boundary
is reported for the collisional (a) and the equilibrium
 (b) scheme as a function of $\tau$.
Measures were taken at the  times shown in figure 4.
\item{Figure 6}: 
Configurations of the system are shown in the left column at different 
values of the shear strain $\gamma t$. In the right column the structure
factor is contour-plotted at the same values of   $\gamma t$.
\item{Figure 7}: 
Evolution of the average domain size in the shear (lower curve) and flow
(upper curve) directions. The slope of $R_x$ is $1.1$.
\end{description}
\newpage
\begin{figure}[h]
\begin{center}
   \includegraphics*[width=0.7\textwidth]{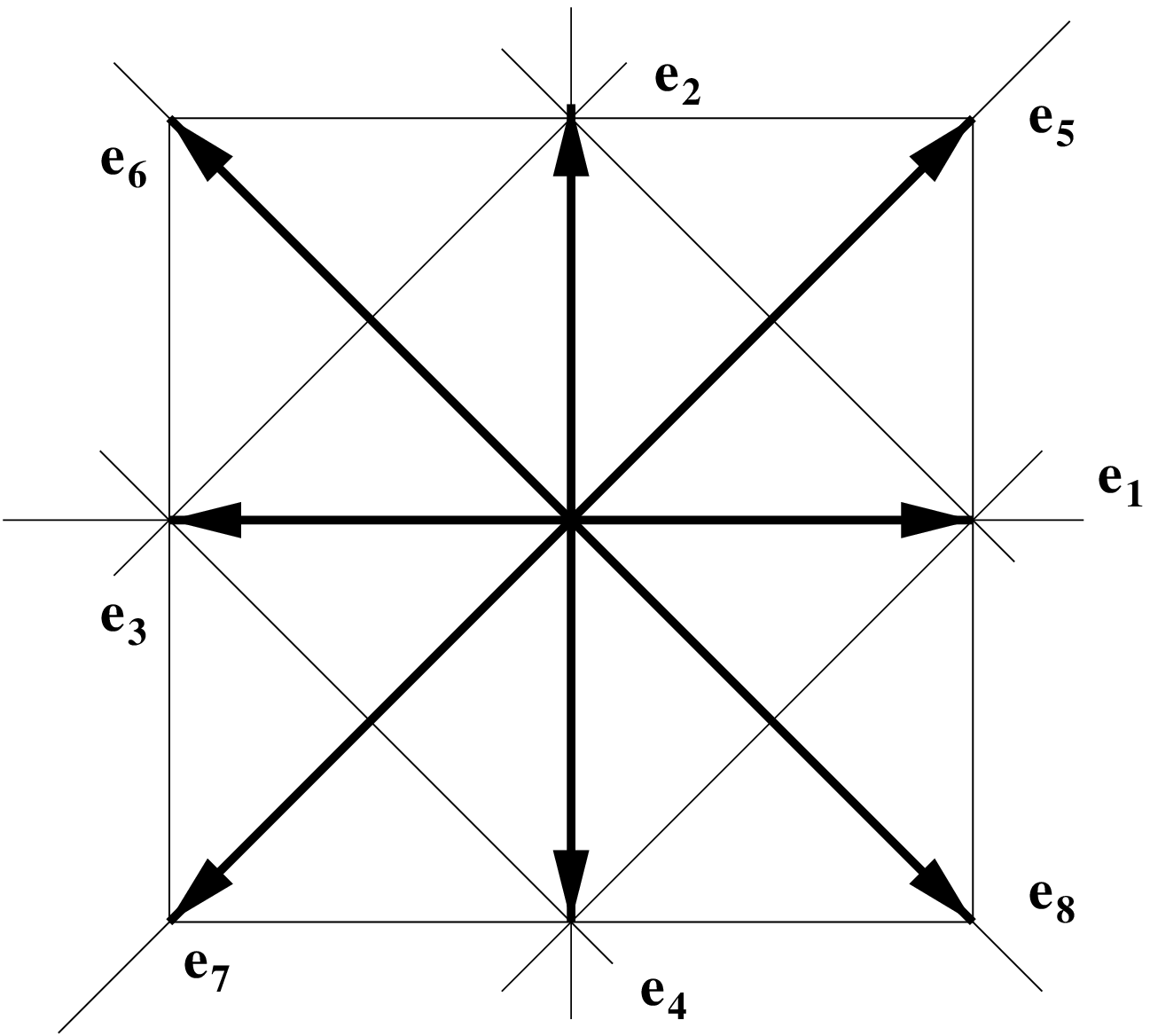}\\*
\end{center}
\label{fig1}
\end{figure}
\begin{center}
Figure 1
\end{center}
\newpage
\begin{figure}[h]
\begin{minipage}{0.45\textwidth}
\centerline{\epsfig{file=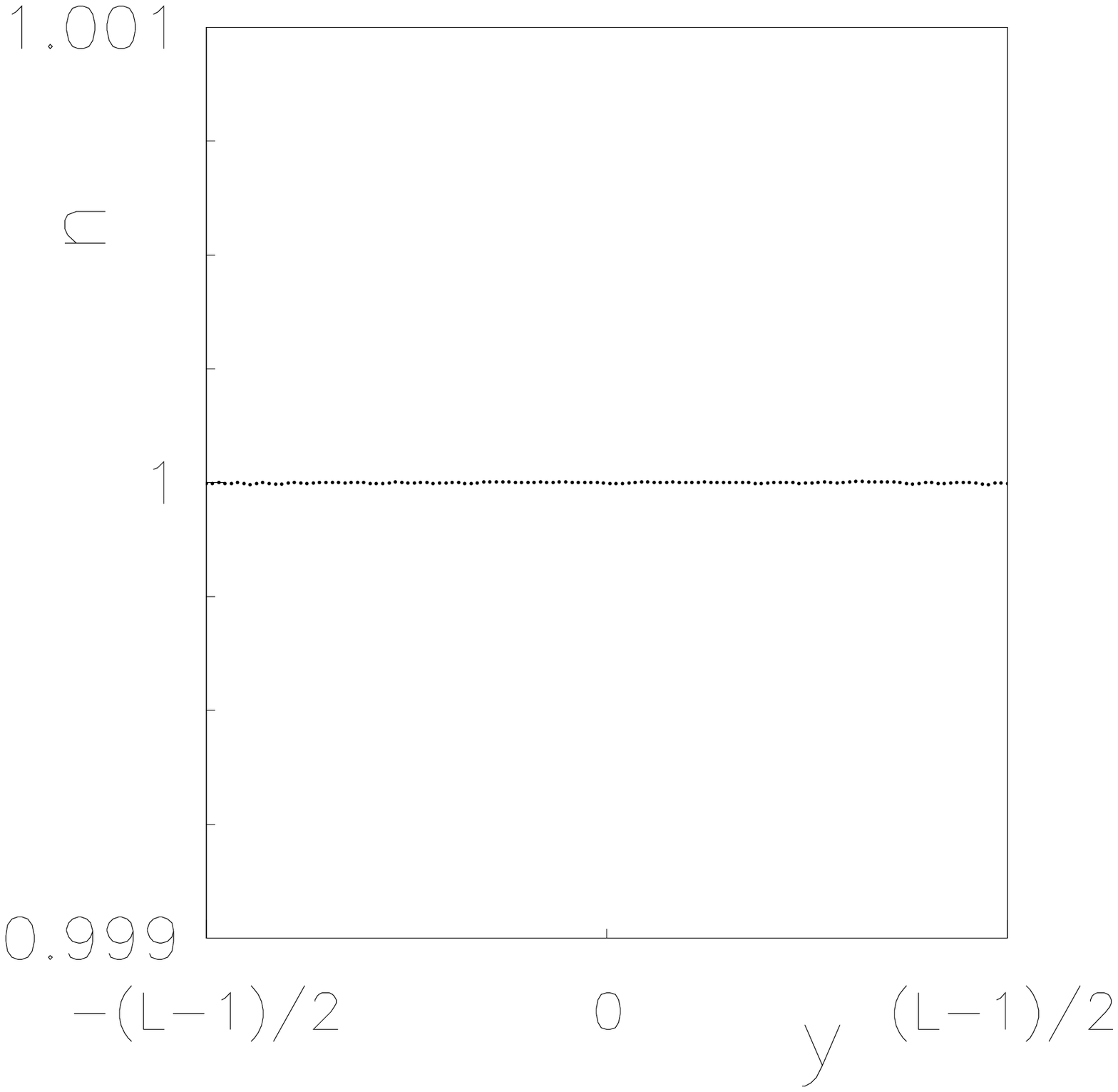,bbllx=17pt,bblly=151pt,bburx=523pt, bbury=660pt,width=0.95\textwidth,clip=}}
\begin{center}
(a)
\end{center}
\end{minipage}
\begin{minipage}{0.45\textwidth}
\centerline{\epsfig{file=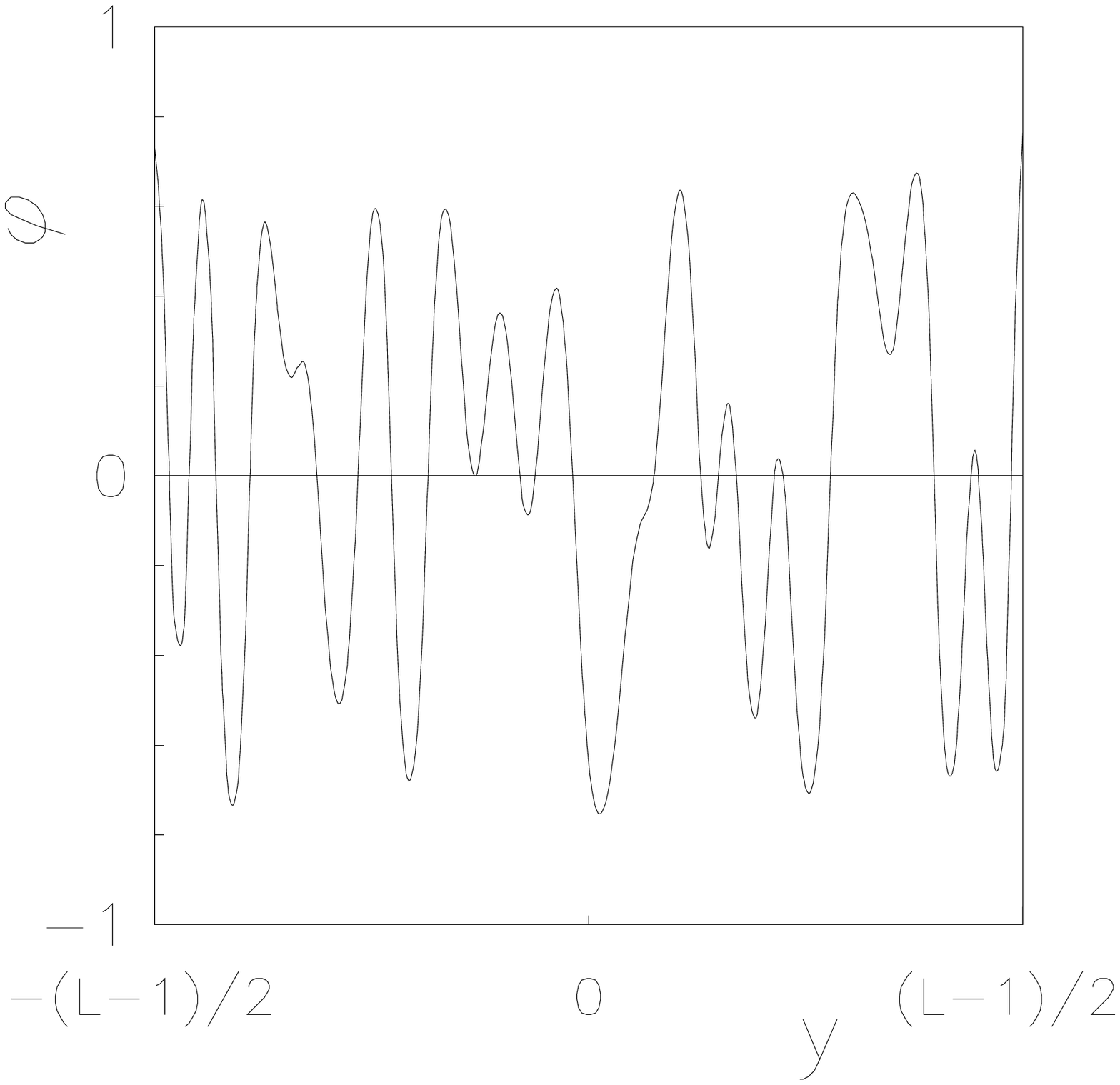,bbllx=17pt,bblly=151pt,bburx=523pt, bbury=660pt,width=0.95\textwidth,clip=}}
\begin{center}
(b)
\end{center}
\end{minipage}
\end{figure}
\begin{center}
Figure 2
\end{center}
\newpage
\begin{figure}[h]
\begin{minipage}{0.8\textwidth}
\centerline{\epsfig{file=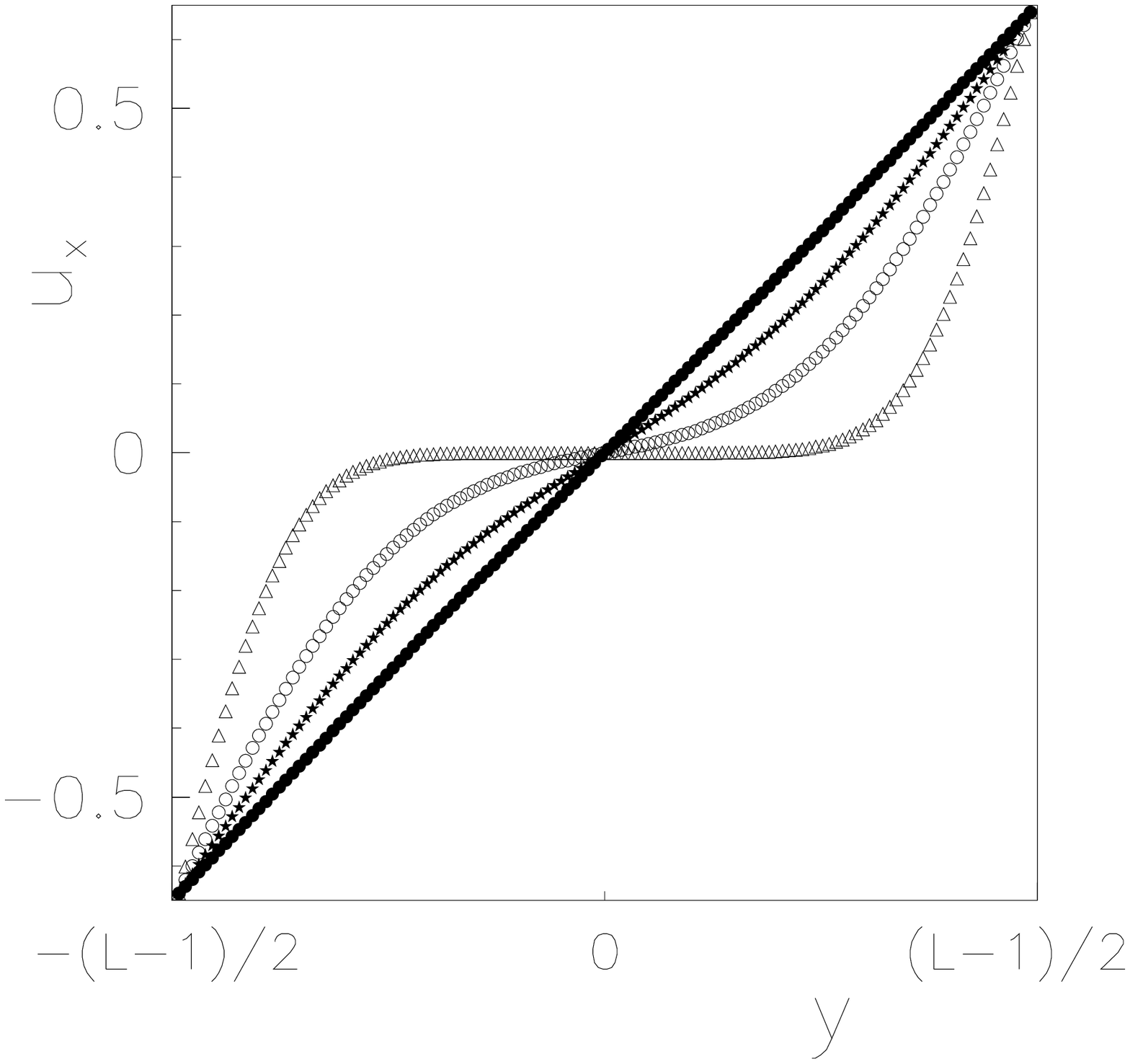,bbllx=17pt,bblly=151pt,bburx=523pt, bbury=660pt,width=0.95\textwidth,clip=}}
\end{minipage}
\end{figure}
\begin{center}
Figure 3
\end{center}
\newpage
\begin{figure}[h]
\begin{minipage}{0.8\textwidth}
\centerline{\epsfig{file=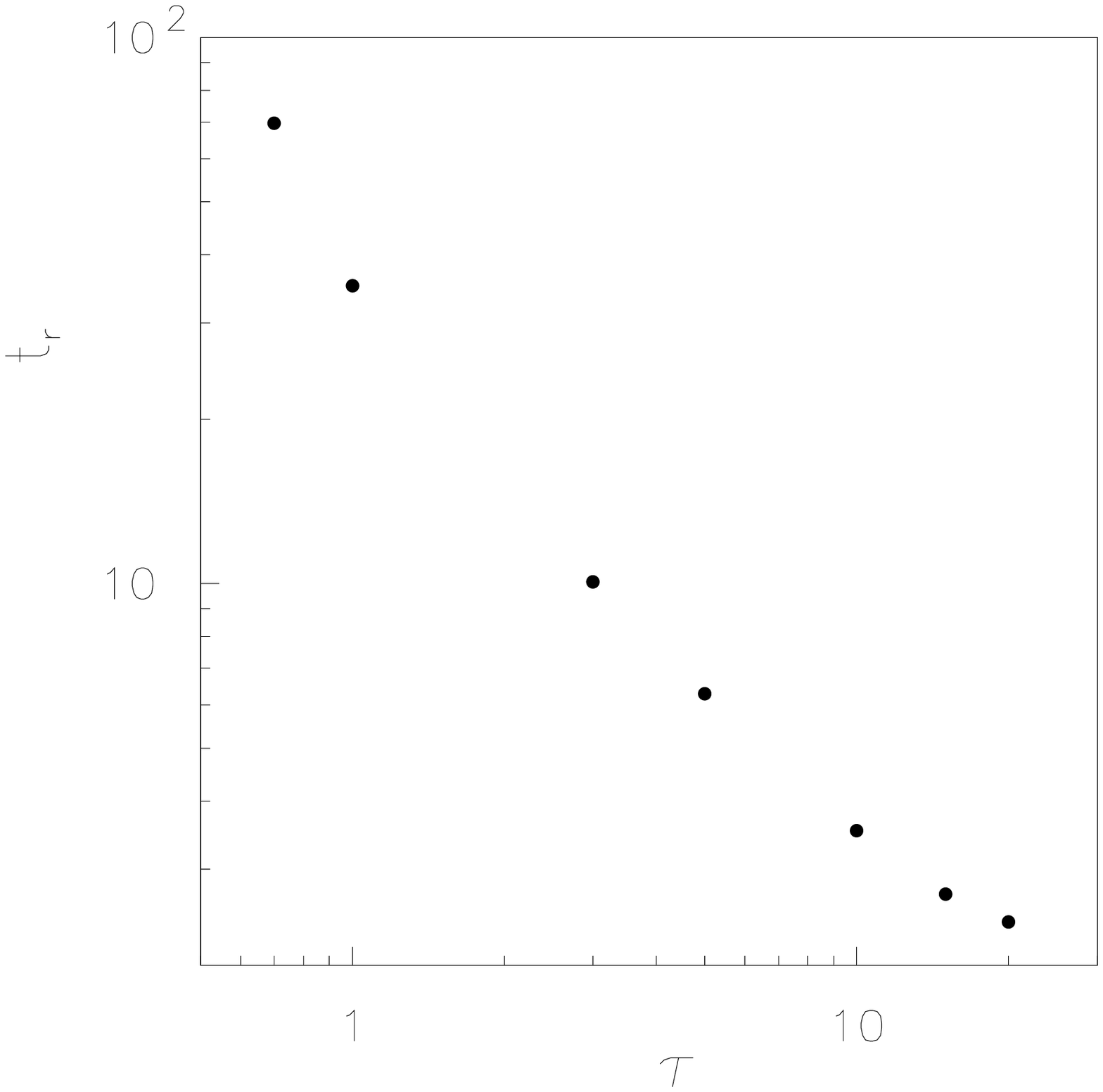,bbllx=22pt,bblly=168pt,bburx=525pt, bbury=662pt,width=0.95\textwidth,clip=}}
\end{minipage}
\end{figure}
\begin{center}
Figure 4
\end{center}
\newpage
\begin{figure}[h]
\begin{minipage}{0.45\textwidth}
\centerline{\epsfig{file=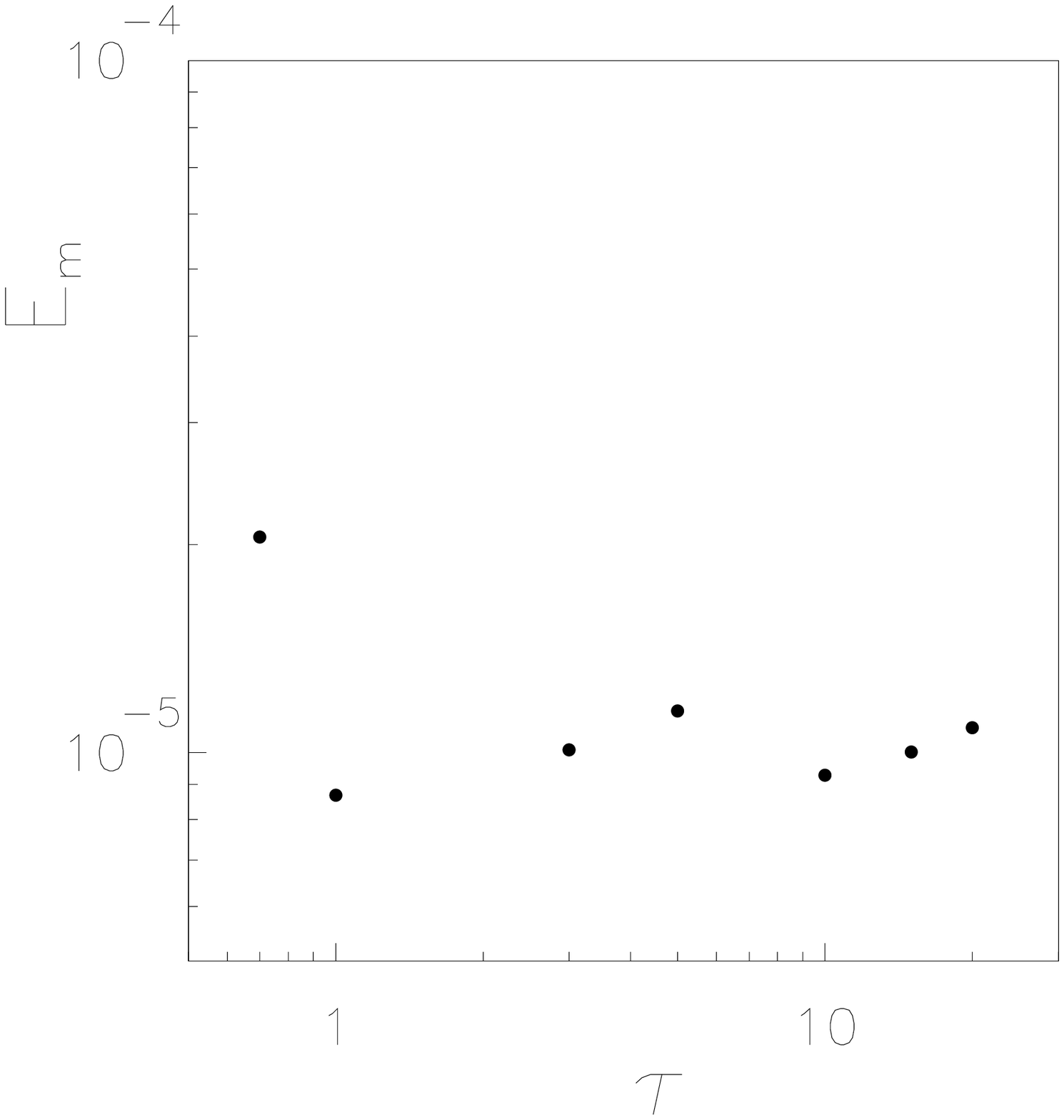,bbllx=27pt,bblly=154pt,bburx=523pt, bbury=646pt,width=0.95\textwidth,clip=}}
\begin{center}
(a)
\end{center}
\end{minipage}
\begin{minipage}{0.45\textwidth}
\centerline{\epsfig{file=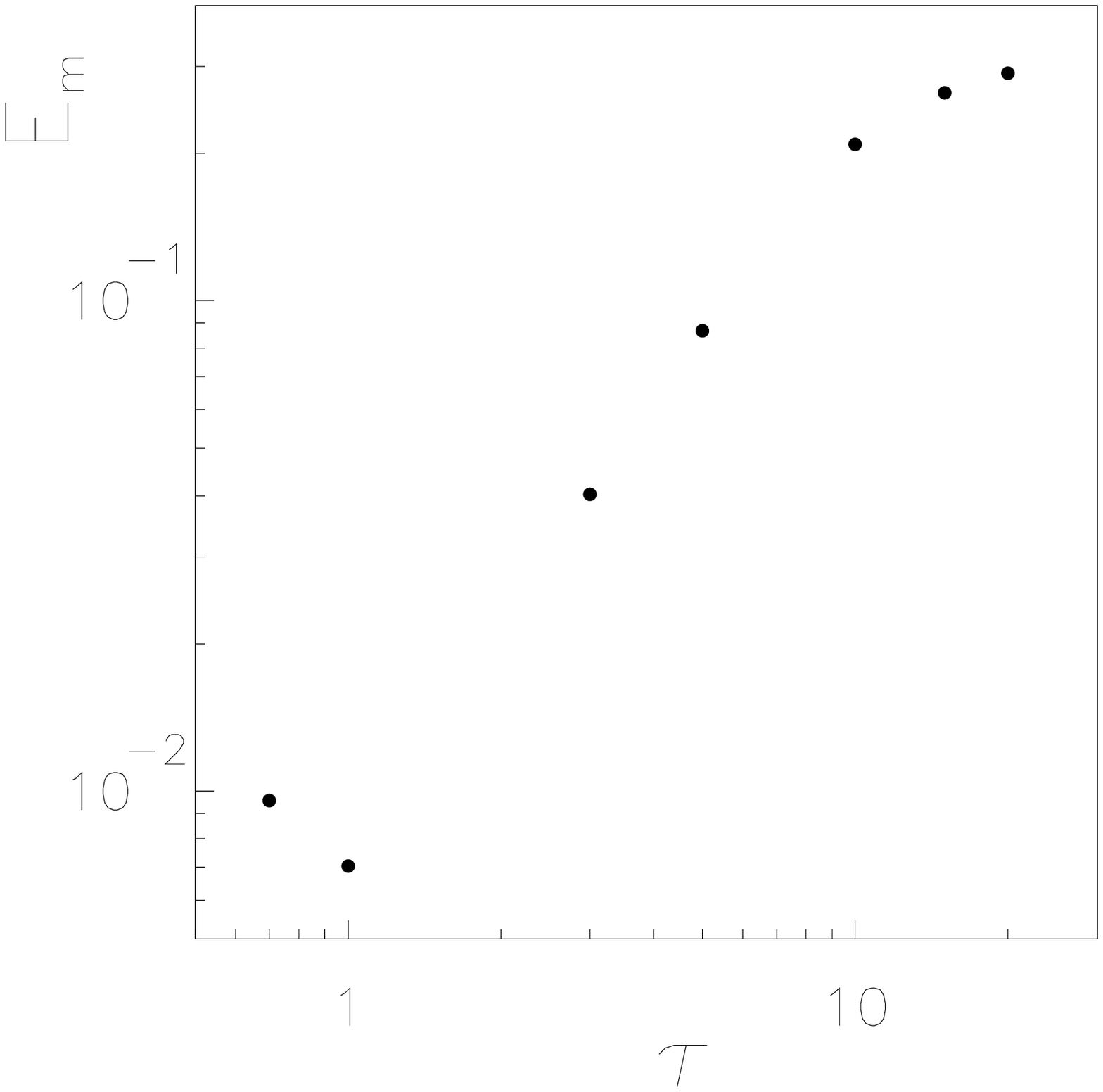,bbllx=26pt,bblly=158pt,bburx=523pt, bbury=646pt,width=0.95\textwidth,clip=}}
\begin{center}
(b)
\end{center}
\end{minipage}
\label{fig5}
\end{figure}
\begin{center}
Figure 5
\end{center}
\newpage
\begin{figure}[t]
\vskip -2.3truecm
\centerline{\epsfig{file=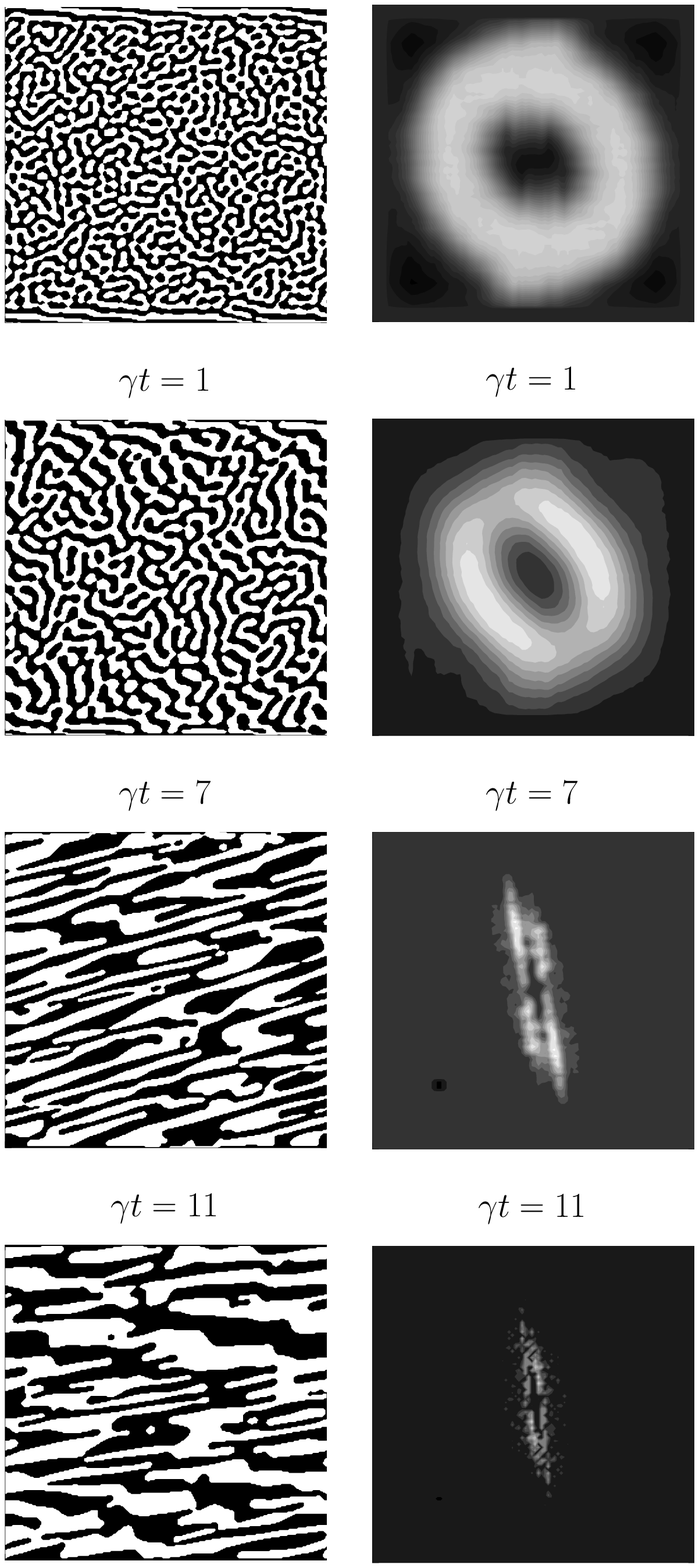,bbllx=69pt,bblly=17pt,bburx=418pt, bbury=825pt,height=0.8\textheight,clip=}}
\label{fig6}
\end{figure}
\begin{center}
Figure 6
\end{center}
\newpage
\begin{figure}[t]
\begin{minipage}{0.8\textwidth}
\centerline{\epsfig{file=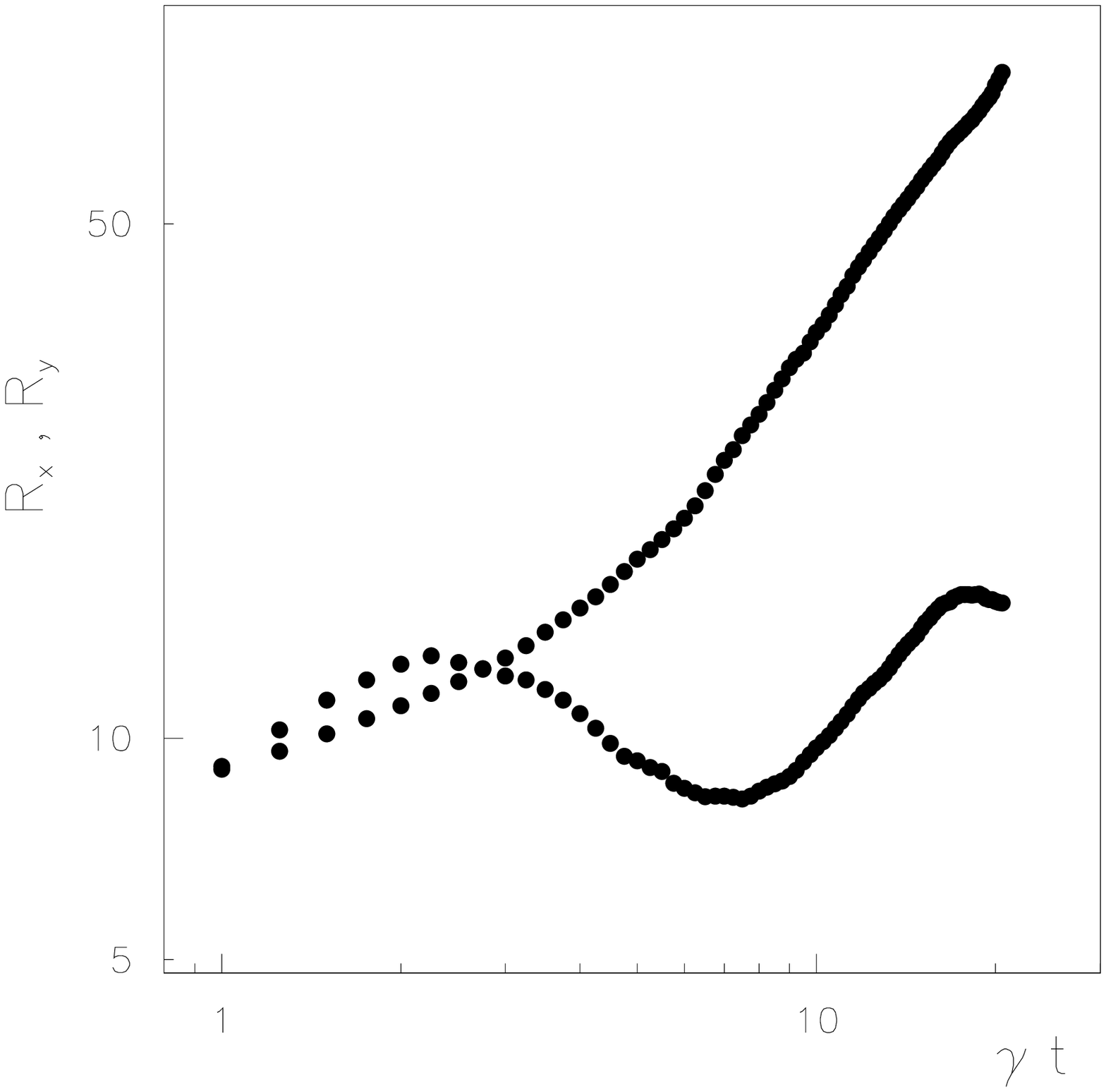,bbllx=14pt,bblly=138pt,bburx=525pt, bbury=650pt,width=0.99\textwidth,clip=}}
\end{minipage}
\label{fig7}
\end{figure}
\begin{center}
Figure 7
\end{center}
\end{document}